\documentclass[a4paper]{proc}
\usepackage{graphicx}
\usepackage{epsfig}
\usepackage[english]{babel}
\usepackage[utf8x]{inputenc}
\usepackage{amsmath}
\usepackage{amssymb}
\usepackage{authblk}
\usepackage{algorithm}
\usepackage{algpseudocode}
\usepackage{hyperref}

\title{"Active Neighbour": A Novel Monitoring Model for Cyber-Physical Systems}
\author{Vasileios Apostolidis-Afentoulis}
\affil{Department of Applied Informatics, School of Information Sciences, University of Macedonia, 156 Egnatia Str., GR-54636 Thessaloniki, Greece}
\affil{\textit {{vapostolidis@uom.edu.gr}}}

\begin{document}

\maketitle

\abstract{Over the past decade, advancements in technology have enabled Cyber-Physical Systems (CPS) to monitor sensor networks through various methodologies. However, these developments have concurrently introduced significant security challenges, necessitating robust protective measures. As a result, securing CPS has become a critical area of research. This paper reviews existing CPS monitoring models and introduces an innovative role-based monitoring model designed to meet contemporary security requirements. The proposed model is implemented within the COOJA simulator of the Contiki OS and evaluated under three distinct security configurations. Preliminary results demonstrate promising outcomes, although further comprehensive testing is ongoing.}


\section{Introduction}
\label{Introduction}
This paper examines the monitoring challenges associated with Cyber-Physical Systems (CPS), an essential domain in research and development (R\&D) \cite{Schoitsch2014}. Ensuring the security of CPS monitoring procedures is paramount for preserving networked computational resources and managing the multi-physical interactions inherent to CPS. Since 2010, the European Research and Industrial Community has identified CPS as a key paradigm for the future of complex systems.

A crucial aspect of CPS monitoring is security, as any vulnerabilities may lead to operational disruptions, particularly in real-time environments. If a monitoring system lacks adequate security measures, fundamental properties such as confidentiality, privacy, and data integrity are at risk \cite{Kane2014}. Effective CPS monitoring requires the capability to aggregate data from multiple network nodes and present it in a structured and meaningful way to system administrators.

\subsection{Monitoring and Specification of CPS}
The effective monitoring of CPS does not solely rely on centralized decision-making and information management centers; instead, it is fundamentally dependent on distributed sensor networks \cite{Shi2011}. The dynamic nature of certain cyber threats limits the effectiveness of traditional management centers in detecting and mitigating attacks in real time, particularly when sensor data is compromised. In scenarios where human intervention is not feasible due to environmental constraints, alternative protective mechanisms must be implemented.

In large-scale CPS deployments, complex interactions exist between semi-autonomous entities, necessitating the application of well-defined software principles and architectural abstractions to facilitate system composition and management.

Over the past 15 years, research efforts have focused on integrating monitoring capabilities into networks and systems. Three primary approaches have emerged: polling-based models, trust management models, and community-based models. Each model employs distinct evaluation techniques for CPS monitoring; however, many fail to incorporate adequate security mechanisms in their communication protocols, which can negatively impact system stability and performance.

To address these limitations, we propose an innovative role-based monitoring model known as the "Active Neighbour" (AN) model. This model introduces two novel security-enhancing mechanisms and is structured around a dynamic role hierarchy. By ensuring efficiency, security, and authentication, AN strengthens communication robustness. The dynamic role-assignment feature maintains continuous network operation, while its adaptive and protective capabilities provide enhanced resilience against cyber threats. The integrated security mechanisms within the AN model improve both the safety and speed of communications.

The remainder of this paper is structured as follows. Section \ref{Related-Work} provides an in-depth review of existing CPS monitoring models. Section \ref{Proposed Scheme} presents the "Active Neighbour" model, detailing its architectural design and implementation. Section \ref{Case Study and Obtained Results} discusses a case study along with the results of simulated experiments. Finally, Section \ref{Conclusions} concludes the paper and outlines directions for future research.

\section{Related Work}
\label{Related-Work}
The scientific community has proposed several models for monitoring CPS to ensure their proper operation. These models can be broadly categorized into three main types:

\begin{enumerate}
 \item \textbf{Polling Models}: They periodically query the network's monitoring services to assess the status of individual nodes.
 \item \textbf{Trust Management Models}: They facilitate the exchange and sharing of trust and reputation information based on nodes' behavior within the network, such as their communication and cooperation with other nodes.
 \item \textbf{Community Models}: They assign roles to participating nodes, each governed by distinct policies that define permitted and prohibited actions. This approach enables the formation of mission-oriented, dynamically self-managing communities tailored to application-specific requirements.
\end{enumerate}

\subsection{Polling Models}
Big Brother (BB) \cite{MacGuire1997} was the first monitoring system to use the Internet as an interconnection point for users. This system allows users to gain a holistic view of network operations in a simplified manner. BB employs a client-server model and incorporates both push and pull data transmission mechanisms. Clients periodically send status updates to the monitoring server, while the network is assessed by polling all available monitoring services. Results are displayed in a central management unit, with each reference including an expiration date to indicate its validity. Typically, network monitoring models require dedicated software installation, limiting administrative flexibility. A distinguishing feature of BB is its web-based interface, which provides real-time status visualization accessible from any internet-connected device.

Zenoss \cite{Badger2008} is an open-source platform that utilizes the Simple Network Management Protocol (SNMP) \cite{Fedor1990,Stallings1993} to monitor networks, servers, applications, and services. The primary advantage of Zenoss is its open architecture, which allows for extensive customization. Since SNMP serves as the underlying protocol, the notification mechanism supports both polling and traps. Additionally, Zenoss enables the storage of historical data, facilitating performance analysis and insight into past operations.

\subsection{Trust Management Models}
The Agent-Based Trust and Reputation Management (ATRM) model \cite{Boukerch2007} is designed for wireless sensor networks (WSNs), where trust and reputation are managed locally with minimal overhead in terms of messaging and time delays. However, as mobile agents traverse the network and execute on remote nodes, they must be initiated by trusted entities.

Another agent-based trust model for WSNs \cite{Chen2007} employs a watchdog mechanism to observe node behavior and broadcast trust ratings. Sensor nodes receive these ratings from agent nodes, which are responsible for trust information collection, processing, and dissemination.

The reputation-based scheme DRBTS \cite{Srinivasan2006} enables beacon nodes (BNs) to monitor each other and provide trustworthiness information to sensor nodes (SNs) based on a voting mechanism. To validate a BN's information, a sensor node must receive trust votes from at least half of its common neighbors.

BTRM-WSN \cite{Marmol2011} is a bio-inspired trust and reputation model for WSNs that identifies the most reliable path to the most reputable node offering a particular service. Each node maintains a trace—analogous to pheromones in biological systems—for each neighbor.

The CONFIDANT model \cite{Buchegger2002} extends routing protocols with a reputation-based mechanism to isolate nodes that deviate from expected behavior. Each node monitors the behavior of its next-hop neighbor, with trust relationships and routing decisions based on observed, experienced, or reported routing and forwarding actions.

The SORI scheme \cite{He2004} incentivizes packet forwarding and discourages selfish behavior. A node's reputation is quantified using objective measures, and reputation propagation is secured via a hash chain-based authentication mechanism.

\subsection{Community Models}
The abstract model \cite{Schaeffer2008} for policy-based collaboration relies on task-oriented roles within sensor communities. Each role is associated with two policy classes:
\begin{enumerate}
 \item \textbf{Obligations}: Define adaptive actions that a role must perform in response to specific events (e.g., device failures or contextual changes).
 \item \textbf{Authorizations}: Define permitted or prohibited actions between roles. The framework supports a unified specification mechanism for both application-specific and management roles.
\end{enumerate}

Node role assignment is constrained to ensure integrity, confidentiality, and availability. Cardinality constraints specify the minimum and maximum number of nodes per role, while separation-of-duty constraints prevent conflicting role assignments. This role-based model explicitly distributes management responsibilities across community participants, enhancing scalability and robustness by avoiding single points of failure.

\subsection{Comparison of the Existing Models}
Both Zenoss and BB have been extensively used in industry. Due to their wide applicability and ease of installation, they remain integral to many monitoring frameworks \cite{Gupta2015}. Comparative analysis indicates that Zenoss is superior to BB for several reasons. Notably, Zenoss supports IPv6, which is crucial for managing the vast number of devices in a CPS. Furthermore, Zenoss leverages SNMP features to collect remote device information, making it a robust monitoring solution.

Trust management models play a crucial role in safeguarding large-scale CPSs against malicious attacks. These mechanisms foster cooperation among distributed entities, identify unreliable sources, and aid decision-making. The ATRM model mitigates network congestion by optimizing communication latency. DRBTS is well-suited for dense networks and can adapt to specific security needs. BTRM-WSN meets CPS requirements for security, scalability, and resource efficiency, consistently delivering robust and accurate results in dynamic, variable, and static networks. Even when malicious servers exceed 60\%, reliable server selection remains above 90\%, with a standard deviation of approximately 7.5\%. The CONFIDANT model effectively detects, warns, and responds to data forwarding and routing attacks. While introducing slight communication delays, it remains scalable and performs well even when 60\% of CPS nodes are compromised. The SORI model efficiently identifies and addresses problematic nodes using three key mechanisms:

\begin{enumerate}
 \item Objective quantification of node reputation.
 \item Secure reputation propagation via an authentication hash chain.
 \item Localized reputation propagation to minimize communication delays.
\end{enumerate}

Community models facilitate autonomous system collaboration through dynamic community formation. Their structural models, task distribution, and communication frameworks offer promising avenues for designing large-scale CPSs. Architectural models abstract sensor functions, enhancing interoperability and reuse. Collaborative CPS architectures can retrospectively implement new structures, streamlining large-scale CPS composition.

While polling, trust management, and community models offer distinct advantages, each has limitations. Polling models are vulnerable to node cloning, leading to false results. Trust management models enhance security but remain insufficient in addressing emerging threats. Community models rely on predefined hierarchies, limiting adaptability. Given evolving security threats \cite{Nur2016,Miao2017} and the emergence of new vulnerabilities \cite{Mitchell2016,Lu2016}, more effective and secure models are required to meet contemporary CPS security demands.

\section{Proposed Scheme – The “Active Neighbour”}
\label{Proposed Scheme}
\subsection{Key Features of the Model}

The proposed model, “Active Neighbour,” introduces dynamic role-assignment procedures to enhance the security of sensor networks and their communications. It builds upon the foundational principles of the “Community Sensor Model based on Roles” while incorporating three critical extensions:

\begin{enumerate}
 \item \textbf{Dynamic Role Assignment:} This feature reinforces self-configuration mechanisms by allowing hierarchical role adaptations based on network conditions and node requirements. For instance, a sensor initially assigned an administrative role may be dynamically downgraded to a lower-level role, thereby mitigating security threats. By ensuring that privileged nodes are not static targets, this mechanism strengthens network resilience against attacks.
 \item \textbf{Time-based One-Time Authentication Algorithm (TOTA):} Unlike the Time-based One-Time Password Algorithm (TOTP) \cite{MRaihi2011}, the TOTA mechanism employs a sequence of rapid authentication challenges based on pre-established network knowledge. This method ensures secure communications both between individual sensors and between sensors and the broader system.
\end{enumerate}

\subsection{Advantages}
The “Active Neighbour” model offers several advantages:

\begin{enumerate}
 \item \textbf{Adaptive Role Assignment:} Sensor roles are dynamically adjusted to align with network requirements, ensuring continuous monitoring while preserving operational integrity.
 \item \textbf{Enhanced Security:} The model proactively defends against malicious attacks through real-time monitoring by “active neighbours.” Threats targeting specific nodes are mitigated via continuous updates to the central management unit.
 \item \textbf{Secure Communication:} The incorporation of TOTA ensures robust authentication mechanisms, reducing vulnerabilities in sensor-to-sensor and sensor-to-system communications.
\end{enumerate}

\subsection{Architectural Design}
\begin{figure}
    \includegraphics[width=\linewidth]{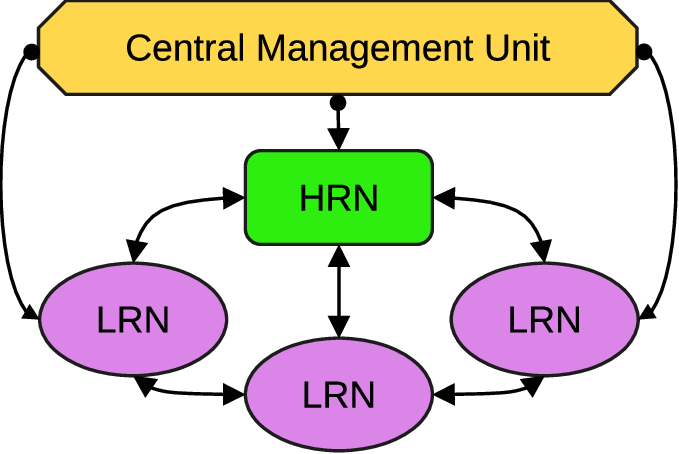}
    \caption{Active Neighbour’s architectural design.}
    \label{ActiveNeighborArchitecturalDesign}
\end{figure}

The architectural framework of the “Active Neighbour” model (Figure \ref{ActiveNeighborArchitecturalDesign}) consists of three primary entities: the Central Management Unit (CMU), High-Rank Nodes (HRNs), and Low-Rank Nodes (LRNs).

\begin{enumerate}
 \item \textbf{Central Management Unit (CMU):} The CMU holds the highest level of authority, responsible for creating and assigning roles, generating cryptographic keys, and maintaining hardware identifiers for network nodes. Once initial assignments are made, the CMU transitions to a supervisory role, intervening only when anomalies are detected.
 \item \textbf{High-Rank Nodes (HRNs):} These nodes perform frequent monitoring tasks, assuming various roles such as administrators, policy enforcers, and authenticity providers. In addition to their supervisory duties, HRNs participate in general network communications.
 \item \textbf{Low-Rank Nodes (LRNs):} Although LRNs primarily execute tasks as directed, they serve a crucial role as the network’s foundational structure. These nodes facilitate communication, process environmental data, and support overall network operations.
\end{enumerate}

The functional responsibilities of HRNs are classified as follows:

\begin{enumerate}
 \item \textbf{Administrator:} Periodically broadcasts status messages, assigns roles to new nodes, and maintains network topology awareness.
 \item \textbf{Policy Applier:} Ensures compliance with network policies by monitoring and reassigning roles as needed. Policy enforcement dictates permissible actions and mandatory procedures for addressing network issues.
 \item \textbf{Authenticity Provider:} Verifies the identity of nodes attempting to join the network, ensuring secure community membership.
\end{enumerate}

All inter-entity communications (as depicted in Figure \ref{ActiveNeighborArchitecturalDesign}) are encrypted and authenticated using Digital Signature Algorithm (DSA) and the IPSec protocol with 256-bit ECCDH (Elliptic Curve Cryptography Diffie–Hellman), providing security equivalent to RSA-3072. The CMU, as the highest authority, can request data streams on demand or in response to communication failures. Informational messages propagate through HRNs and LRNs to maintain network synchronization. HRNs autonomously monitor LRNs and notify the network of any malfunctioning nodes, ensuring operational continuity.

\subsection{Dynamic Role-Assignment Algorithm} In order to dynamically assign roles to the sensors of the CPS’s network, a sequence of steps is followed. This dynamic process not only gives the model its adaptive character but also enhances the security of the underlying communications. The details of this dynamic role-assignment procedure are outlined in Algorithm \ref{alg:dyn_role}. In brief, the nodes are periodically queried by the CMU and, according to their role type and status are treated accordingly. The “queries” are in the form of request/response messages that get exchanged between the communicating nodes. The role assignment procedure is triggered whenever a node is not working properly (for the sake of simplicity, it is currently examined whether it is dropping packets).

Role assignment follows a structured selection process, prioritizing candidates based on network topology. Each node pings its peers, and response times are evaluated to determine the optimal candidate for reassignment. If the selected node becomes unresponsive, the next best candidate is chosen, ensuring robust network operation.

\begin{algorithm*}[ht]
\caption{Active Neighbor's Dynamic Role Assignment in CPS}\label{alg:dyn_role}
\begin{algorithmic}
\State \textbf{Start}
\State Specify the number of sensors that comprise the CPS.
\State Detect the sensor with the highest processing power and assign it an administrator's role.
\State Inspect the sensors' function.
\If{all sensors operate properly}
    \State All roles remain the same
    \State \textbf{Terminate}
\Else
    \State Begin dynamic role assignment process
\EndIf

\State Inspect the function of the sensor with the administrator's role.
\If{administrator sensor operates properly}
    \State Begin inspection of the function of the remaining sensors.
\Else
    \State Assign the administrator's role to the next sensor in hierarchy.
    \State Transmit informational message to the rest of the sensors.
\EndIf

\State Inspect the function of the remaining sensors.
\If{all sensors operate properly}
    \State Move to the next step
\Else
    \State Locate malfunctioning sensor from its neighbors.
    \State Transmit informational message about the change.
\EndIf

\State Re-inspect the operation of the sensors.
\If{CPS network operates properly}
    \State \textbf{Terminate}
\Else
    \State Go back to Step 4
\EndIf
\State \textbf{End}
\end{algorithmic}
\end{algorithm*}

\subsection{Network Instability Scenarios} In cases where communication or monitoring quality is compromised, the “Active Neighbour” model issues specific warnings and informational messages, enabling prompt corrective actions.

\subsubsection{Fire Sensor Notifications}
When a fire sensor experiences any of the following issues:

\begin{enumerate}
    \item Loss of a single data packet triggers a warning.
    \item Loss of three consecutive data packets results in an alert.
\end{enumerate}

Upon detecting three consecutive packet losses, the following steps are taken:

\begin{enumerate}
    \item The network issues an informational message indicating the removal of the fire sensor due to communication failure.
    \item A diagnostic mechanism initiates periodic test data transmissions to the removed sensor at one-minute intervals.
\end{enumerate}

\subsubsection{Administrator Sensor Notifications}
If the administrator sensor experiences:

\begin{enumerate}
    \item A single data packet loss, a warning is generated.
    \item Three consecutive packet losses, an alert is issued, triggering role reassignment to the next suitable sensor in the hierarchy.
\end{enumerate}

Upon three consecutive packet losses, the following steps are executed:

\begin{enumerate}
    \item An informational message is generated, confirming the removal of the administrator sensor due to communication failure.
    \item A notification specifies the newly designated administrator sensor.
    \item The previous administrator sensor undergoes a reassignment process and is designated as a low-rank sensor upon network reentry.
\end{enumerate}

\section{Case Study and Obtained Results}
\label{Case Study and Obtained Results}
\subsection{Experimental Setup}

To evaluate the "Active Neighbour" model, experiments were conducted in a simulation environment. Specifically, the COOJA simulator of Contiki OS v2.7 was utilized \cite{Dunkels2004}. Contiki OS is an open-source operating system designed for resource-constrained devices. The objective was to create a sensor network within a CPS to monitor climate conditions in a forest environment. The experiments were run on a system equipped with an Intel i7 4770 processor and 8 GB of RAM.

To align the hierarchical structure of the architectural model with the experimental setup, the following mapping was established: in the case of the High-Responsibility Node (HRN), an administrator role was assigned to a designated node, whereas Low-Responsibility Nodes (LRNs) were represented by fire sensors.

Three simulation profiles were designed to assess the model under different conditions:

\subsection{Case 1: Plain “Active Neighbour”}
\begin{figure}
    \includegraphics[width=\linewidth]{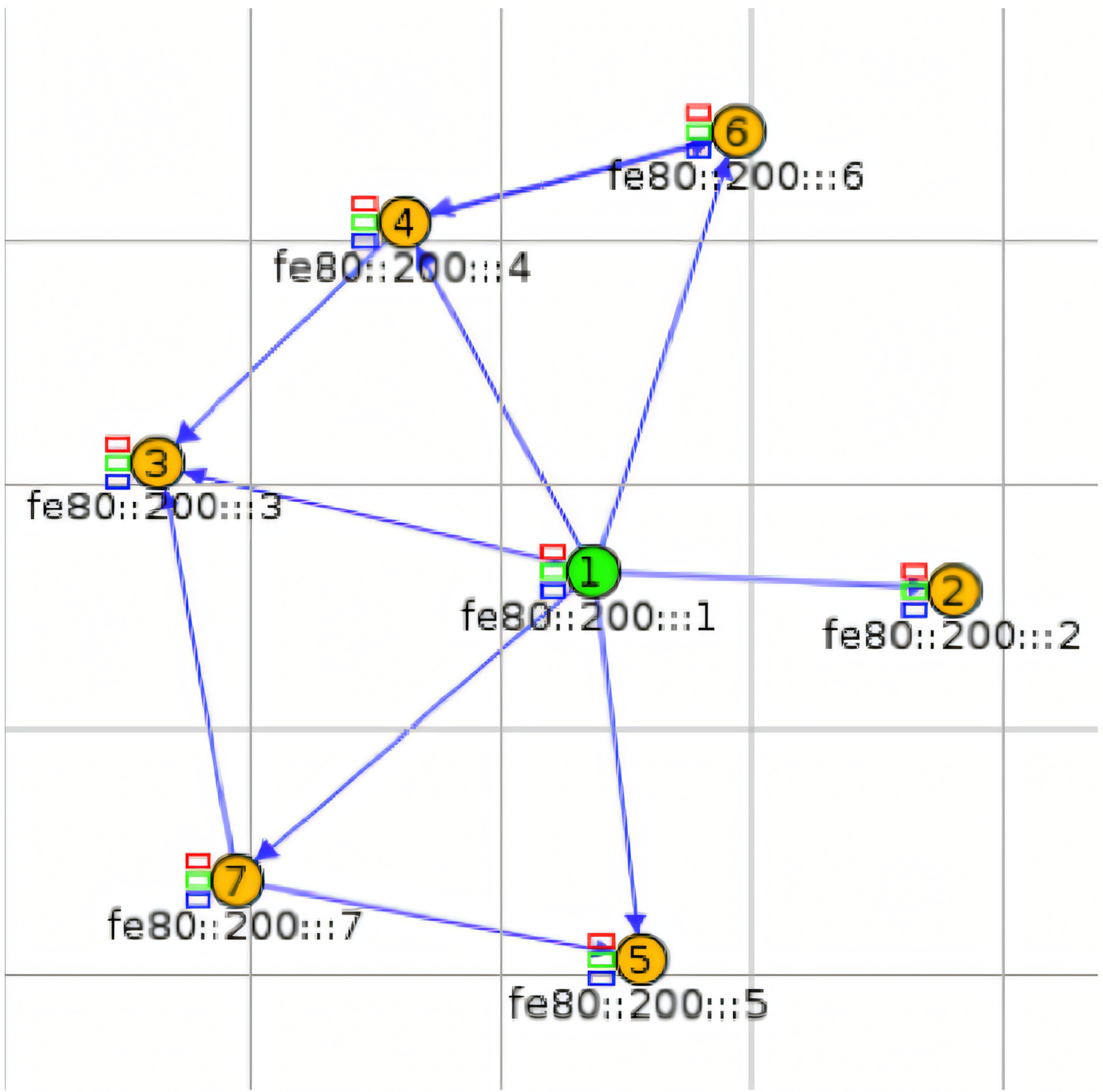}
    \caption{Implementation of the “Active Neighbour” model in Contiki OS.}
    \label{ImplementationOfTheActiveNeighbourModelInContikiOS}
\end{figure}

In this scenario, WiSMote-type nodes were deployed to create the virtual network. Each node is represented as a numbered circle, corresponding to its identifier. As illustrated in Fig. \ref{ImplementationOfTheActiveNeighbourModelInContikiOS}, the node with identifier 1 was assigned the role of administrator (HRN) and is depicted in green. The remaining nodes, with identifiers ranging from 2 to 7, were designated as fire sensors (LRNs) and are shown in yellow. The blue arrows indicate communication links between CPS nodes.

\subsection{Case 2: “Active Neighbour” with Message Authentication}
In this case, the security of the "Active Neighbour" model was enhanced through the implementation of the Digital Signature Algorithm (DSA) to ensure message authenticity. This authentication mechanism provides a security level comparable to RSA-1024.

Upon algorithm initiation, the number of nodes was specified (six in this scenario). The nodes were initialized, and their roles were assigned based on the network topology. The administrator node required high processing power for its assignment. Following this, all network nodes were authorized by the Central Management Unit (CMU), which involved registering the nodes' unique hardware identifiers in a database.

Subsequently, each communication data packet was authenticated using a combination of the nodes’ IDs and the DSA. The system then performed an operational inspection of the sensors. If all nodes functioned correctly, the algorithm terminated, maintaining the assigned roles. Otherwise, the dynamic role assignment process was triggered. During this process, the administrator node was inspected first. If it remained functional, the inspection proceeded to the other nodes. If the administrator node malfunctioned, the role was reassigned to the next hierarchical node, and an informational message was transmitted. The inspection process continued for the remaining nodes, and the overall network functionality was evaluated. If the network operated as expected, the algorithm terminated; otherwise, it reverted to the node inspection phase before concluding.

\subsection{Case 3: “Active Neighbour” with IPSec and Message
Authentication}
In this scenario, security measures were further reinforced by integrating the IPSec protocol alongside DSA. For key establishment, Elliptic Curve Cryptography Diffie–Hellman (ECCDH) was employed, providing a security level equivalent to RSA-3072.

At the onset, the number of nodes was set to seven. Nodes were initialized and assigned roles according to the network topology, with the administrator node requiring high processing power. As in the previous case, all network nodes underwent authorization by the CMU, including the registration of their unique hardware IDs. Each communication data packet was then authenticated and encrypted using a combination of the nodes’ IDs, DSA, and IPSec.

The subsequent steps involved sensor functionality verification. If all nodes operated correctly, the algorithm terminated while maintaining assigned roles. Otherwise, the dynamic role assignment process was activated, beginning with an inspection of the administrator node. If the administrator node was functional, the process proceeded to inspect the remaining nodes. If the administrator node failed, its role was reassigned to the next hierarchical node, accompanied by an informational message. The overall network functionality was then assessed. If the network functioned correctly, the algorithm terminated; otherwise, it reverted to the node inspection phase before completion.

\subsection{Evaluation}
\begin{figure}
    \includegraphics[width=\linewidth]{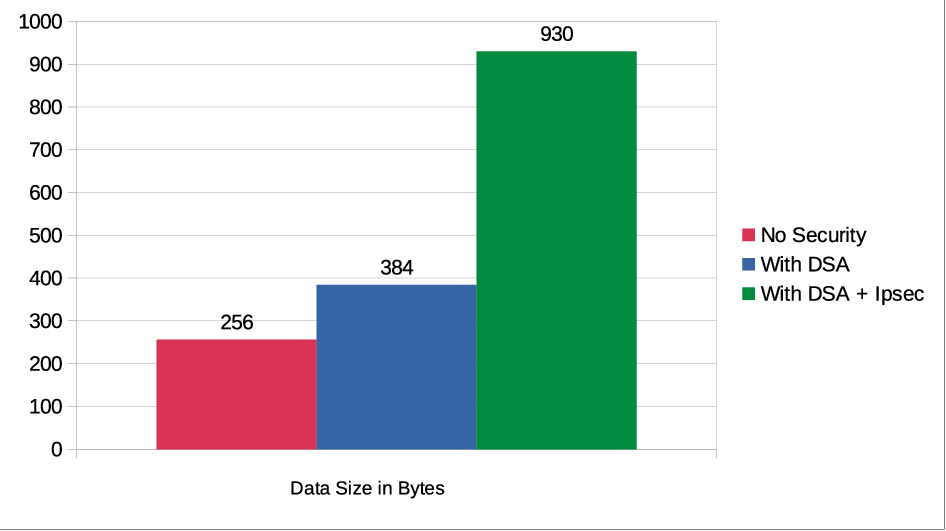}
    \caption{Comparison of the monitoring cases.}
    \label{ComparisonOfTheMonitoringCases}
\end{figure}

Figure \ref{ComparisonOfTheMonitoringCases} summarizes the experimental results. The findings indicate that utilizing DSA alone for message authentication introduces minimal data overhead to network communications. This approach is suitable for scenarios where confidentiality is not a primary concern. Conversely, the combined use of IPSec and DSA significantly increases data overhead, nearly quadrupling the overhead compared to the plain model and tripling it relative to the DSA-only configuration.

\section{Conclusions}
\label{Conclusions}
This study introduced an innovative monitoring model tailored for Cyber-Physical Systems. The primary feature of the proposed model is its capability to dynamically assign roles to CPS nodes, enabling the formation of autonomous CPS communities. Ongoing and future work involves an extensive evaluation of the model, focusing on computational and data overhead, as well as energy consumption, through the analysis of various performance parameters.

\bibliographystyle{plain}
\bibliography{ActiveNeighbourANovelMonitoringModelforCyberPhysicalSystems}

\end{document}